\documentclass[aps,prx,superscriptaddress,twocolumn,showpacs,floatfix]{revtex4-1}

\pdfoutput=1

\usepackage{amsbsy}
\usepackage{amsmath}
\usepackage{amssymb}
\usepackage{amsfonts}
\usepackage{bm}
\usepackage{yfonts}

\usepackage{graphicx}
\usepackage{pst-all, pst-solides3d}

\newcommand{\cohosub}[1]{\scalebox{0.7}{\textswab{#1}}}
\newcommand{\coho}[1]{\textswab{#1}}
\newcommand{\U}[1]{\textrm{U}}

\begin{document}

\date{\today}

\title{Topological Enrichment of Luttinger's Theorem}

\author{Parsa Bonderson}
\affiliation{Station Q, Microsoft Research, Santa Barbara, California 93106-6105, USA}

\author{Meng Cheng}
\affiliation{Station Q, Microsoft Research, Santa Barbara, California 93106-6105, USA}

\author{Kaushal Patel}
\affiliation{Department of Physics, University of California, Santa Barbara, California 93106, USA}

\author{Eugeniu Plamadeala}
\affiliation{Department of Physics, University of California, Santa Barbara, California 93106, USA}

\begin{abstract}
We establish a generalization of Luttinger's theorem that applies to fractionalized Fermi liquids, i.e. Fermi liquids coexisting with symmetry enriched topological order. We find that, in the linear relation between the Fermi volume and the density of fermions, the contribution of the density is changed by the filling fraction associated with the topologically ordered sector, which is determined by how the symmetries fractionalize. Conversely, this places constraints on the allowed symmetry enriched topological orders that can manifest in a fractionalized Fermi liquid with a given Fermi volume and density of fermions.
\end{abstract}


\maketitle


\section{Introduction}

Free electrons in a translationally invariant system form a Fermi sea. Interacting electrons may be described by Landau's Fermi liquid theory, which, when applicable, asserts that interactions between electrons do not qualitatively modify the free electron picture, at most dressing the electrons as quasiparticles, which are fermions with renormalized quantities, such as mass. In particular, the Fermi volume $V_F$ of these emergent quasiparticles is precisely determined by the filling fraction $\nu$ of the underlying electrons per unit cell:
\begin{equation}
\nu=\frac{V_F}{(2\pi)^D} \mod 1,
\end{equation}
where the relation holds modulo an integer, which physically represents the number of filled bands. This relation, fixing the Fermi volume for a specified electron density, is the content of Luttinger's theorem~\cite{luttinger60}, which is a rare example of an exact result for an interacting system.

While Luttinger's theorem was originally proved perterbatively, it was later recast as a ``quantization'' condition for Fermi liquids by Oshikawa~\cite{oshikawa00}, who proved it non-perturbatively by drawing inspiration from Laughlin's flux threading argument~\cite{laughlin81, tao84, tao86} and Lieb, Schultz, and Mattis's variational argument~\cite{lieb61}. Later, it was found that Luttinger's theorem may require modification for a fractionalized Fermi liquid, i.e. a Fermi liquid that is accompanied by symmetry enriched topological (SET) order, and that this modification, at least for some simple cases, may be understood by generalizing Oshikawa's arguments~\cite{senthil03, senthil04, paramekanti04}.

In this paper, we apply Oshikawa's arguments to 2D systems with general SET order~\cite{barkeshli14}. By studying the interplay between symmetries, topological order, and the Fermi sea, we derive a topologically enriched generalization of Luttinger's theorem for fractionalized Fermi liquids:
\begin{equation}
\label{eq:TELutt}
    \nu-\nu_{\text{topo}}=\frac{V_F}{(2\pi)^2} \mod 1,
\end{equation}
where, assuming that the underlying degrees of freedom effectively decouple into an SET sector and a Fermi liquid sector, $\nu_{\text{topo}}$ is the filling fraction of the SET sector. For 2D systems, there is a precise general definition of $\nu_{\text{topo}}$; it is the $\textrm{U}(1)$ charge of the background anyonic flux per unit cell that is specified by the SET order~\cite{cheng15}, as we will describe. In higher dimensions, we expect a similar definition (and verify it for specific 3D examples), but a general formalism of higher dimensional topological and SET order is currently lacking. Our result reaffirms the intuition that the underlying degrees of freedom that topologically order should not contribute to the Fermi volume.

A consequence of the topologically enriched Luttinger's theorem is that experimental observation of a Fermi volume that deviates from that of an ordinary Fermi liquid may point to the existence of a fractionalized Fermi liquid phase. Moreover, the SET order that is allowed for a given deviation is constrained by the corresponding value of $\nu_{\text{topo}}$.

Our paper is organized as follows. In Sec.~\ref{sec:oshikawasargument}, we review Oshikawa's proof of Luttinger's theorem. In Sec.~\ref{sec:symfrac}, we review 2D symmetry fractionalization, focusing on $\textrm{U}(1)$ and translational symmetries, and show that flux threading argument places a constraint on which SET phases are allowed at some given filling. In Sec.~\ref{sec:ffl}, we derive the topologically enriched version of Luttinger's theorem for a general 2D fractionalized Fermi liquid, apply it to the $\mathbb{Z}_2$ fractionalized Fermi liquid (FL*), and examine some examples of 3D $\mathbb{Z}_2$ fractionalized Fermi liquids. Finally, in Sec.~\ref{sec:discussion}, we discuss further possible applications and generalizations of our work.

\section{Oshikawa's Argument}
\label{sec:oshikawasargument}

In essence, Oshikawa's argument involves starting with a periodic system in its ground state, adiabatically inserting a flux along one of the directions, applying a large gauge transformation to remove the flux, and finally comparing the resulting state with the original state in order to derive constraints for the system. This yields the commensurability condition if the system is gapped~\cite{oshikawa99}, and Luttinger's theorem if the system is gapless with a Fermi surface of charged quasiparticles~\cite{oshikawa00}. We review these arguments in more detail.

We consider a $D$ dimensional periodic system of size $L_1 \times \dots \times L_D$ with a global $\textrm{U}(1)$ symmetry and a corresponding filling fraction, specifying the density per unit cell $\nu=p/q$ for some coprime integers $p$ and $q$. We assume the system is described by a translationally invariant Hamiltonian $\bm{H}(0)$ and is in a ground state $|\Psi(0)\rangle$. Let the state $|\Psi(0)\rangle$ be an eigenstate of the translation operator $\bm{R}_{T_1}$ with eigenvalue $e^{i P_1(0)}$, i.e. it has momentum $P_1(0)$. (We set $\hbar=e=1$.)
\begin{figure}[t!]
    \centering
    \includegraphics{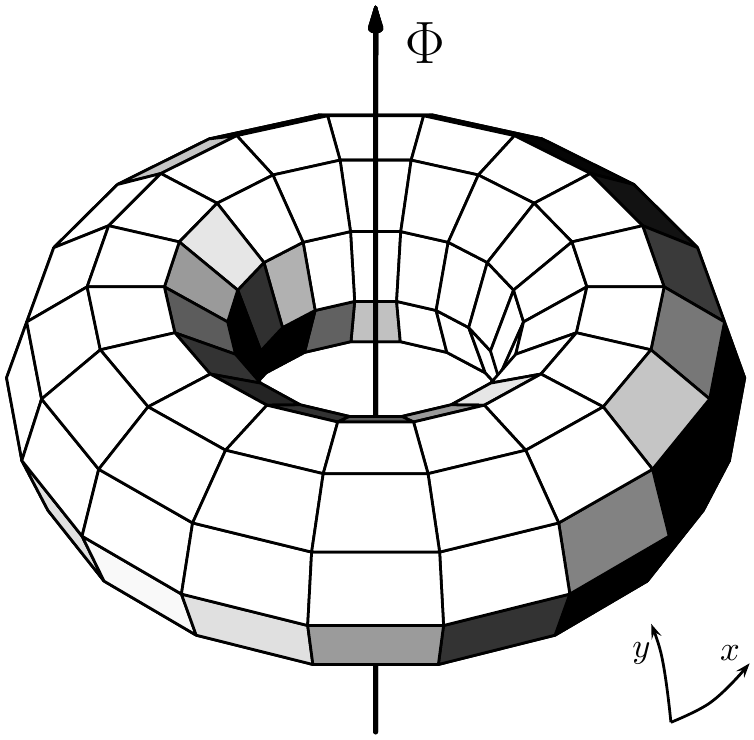}
\caption{A 2D periodic lattice depicted as a torus. A flux inserted to induce a uniform electric field in the $x$-direction can be thought of as threading the handle of the torus.}
    \label{fig:fluxtorus}
\end{figure}

Next, we consider adiabatically inserting a $\textrm{U}(1)$ flux so that a uniform electric field is induced along the $\hat{x}_{1}$-direction, for which the Hamiltonian is $\bm{H}(\Phi)$ and adiabatic path of ground states is given by $|\Psi(\Phi)\rangle$. In the 2D case, where the periodic system is effectively a torus, the flux can be thought of as threading the handle of the torus, as illustrated in Fig.~\ref{fig:fluxtorus}. Since inserting a $2\pi$ flux returns the system to the same point in configuration space, the spectra of $\bm{H}(2\pi)$ and $\bm{H}(0)$ are identical, and there exists a large gauge transformation $\bm{\mathcal{G}}$ that removes the flux: $\bm{\mathcal{G}} \bm{H}(2\pi) \bm{\mathcal{G}}^{-1}=\bm{H}(0)$. Therefore, $\bm{\mathcal{G}}|\Psi(2\pi)\rangle$ must be an eigenstate of $\bm{H}(0)$. Also, since $[\bm{R}_{T_1},\bm{H}(\Phi)]=0$ throughout the flux threading process, $|\Psi(2\pi)\rangle$ has momentum $P_1(0)$, and since $\bm{\mathcal{G}}\bm{R}_{T_1}\bm{\mathcal{G}}^{-1}=e^{i 2\pi \nu L_2\cdots L_D} \bm{R}_{T_1}$, the state $\bm{\mathcal{G}}|\Psi(2\pi)\rangle$ has momentum $P_1(0)+ 2\pi \nu L_2\cdots L_D \mod 2\pi$.

If the system is gapped and remains gapped throughout the flux threading process, then the adiabatic theorem guarantees that $\bm{\mathcal{G}}|\Psi(2\pi)\rangle$ is a ground state of $\bm{H}(0)$ \footnote{For subtleties regarding adiabatic flux insertion and quasi-adiabatic evolution of gapped, degenerate Hamiltonians, see Ref.~\onlinecite{hastings05}.}. By choosing arbitrary integers $L_2$, \dots, $L_D$ that are coprime with $q$, we find $q$ degenerate ground states with different momenta. In the absence of topological order, these degenerate ground states must be the result of spontaneous translational symmetry breaking. Their period in the $\hat{x}_{1}$-direction must be an integer multiple of $q$, and therefore the new unit cell, which is the original unit cell enlarged by a factor of $q$, has an integer filling fraction. This is Oshikawa's commensurability condition~\cite{oshikawa99}, which was later rigorously proven for 2D systems by Hastings~\cite{hastings04}. In the presence of topological order, translational symmetry need not be spontaneously broken, since topological phases can have translationally-invariant, degenerate ground states on a torus. In 2D, however, only certain topological orders can coexist with $\textrm{U}(1)$ and translational symmetry for a given $\nu$, as we will explain in Sec.~\ref{sec:fluxthreading}. In the rest of this paper, we assume there is no spontaneous symmetry breaking in the system.

If the system is gapless, then $\bm{\mathcal{G}}|\Psi(2\pi)\rangle$ is no longer necessarily a ground state. However, it is still true that its momentum is shifted by $2\pi \nu L_2 \cdots L_D$, and this shift can be compared with the momentum shift of the emergent degrees of freedom. For example, if the system is a Fermi liquid of charge 1 quasiparticles, then threading the flux applies a Galilean boost to the Fermi sea, shifting the momentum of each of the $N_F$ quasiparticles by $2\pi/L_1$. Equating the two momentum shifts yields the constraint
\begin{equation}
\nu= \frac{ V_F}{(2\pi)^D } + \frac{ n}{ L_2\cdots L_D},
\end{equation}
for some integer $n$, where $V_F\equiv(2\pi)^D N_F/L_1 \cdots L_D$ is the Fermi volume. Similarly, inserting flux along the other directions yields more constraints on $\nu$, which are compatible for coprime integers $L_1$, \dots, $L_D$ iff
\begin{equation} \label{eq:luttinger}
    \nu=\frac{V_F}{(2\pi)^D} \mod 1.
\end{equation}
This is Oshikawa's derivation of Luttinger's theorem~\cite{oshikawa00}. In Appendix~\ref{ap:kondo}, we provide a derivation of Luttinger's theorem for the 2D Kondo model, under the assumption that it is in a Fermi liquid phase.

\section{Symmetry Fractionalization}
\label{sec:symfrac}

A symmetric system with topological order can manifest distinct SET phases, which cannot be adiabatically connected to each other while respecting the symmetry~\cite{barkeshli14}. A distinguishing signature of these phases is symmetry fractionalization~\cite{essin13,barkeshli14}, a phenomenon that allows quasiparticles to carry fractionalized quantum numbers of the symmetry. For example, $\textrm{U}(1)$ fractionalization leads to quasiparticles with fractional charge~\cite{tsui82, laughlin83}, while translational symmetry fractionalization leads to a nontrivial background anyonic flux in the system~\cite{zaletel15,cheng15}. (Both of these examples will be described in more detail.)

In general, symmetry fractionalization in a 2D topologically ordered phase is classified by the cohomology group $H^2_{\rho}(G,\mathcal{A})$, where $G$ is the symmetry group, $\mathcal{A}$ is the group of Abelian anyons under fusion, and $\rho$ is the symmetry action, which may permute anyon types. We first review the derivation of this classification and, in doing so, introduce relevant notation and concepts. (See Ref.~\onlinecite{barkeshli14} for more details.) We will assume that the topological order is bosonic and that symmetries are unitary. We also focus on the case where the symmetry action $\rho$ does not permute anyon types, which must be the case for symmetries described by a continuous and connected group, such as $\textrm{U}(1)$.

\subsection{Review of On-Site Symmetry Fractionalization}

Consider a symmetric 2D system in a topological phase with symmetry group $G$, whose elements $\mathbf{g}$ act linearly on the Hilbert space by the unitary on-site operators $\bm{R}_\mathbf{g} = \prod_{k \in I} \bm{R}_{\mathbf{g}}^{(k)}$. Let $|\Psi_{\{a_1,\dots,a_n\}}\rangle$ be a state with $n$ quasiparticles carrying topological charges $a_1, \dots, a_n$, respectively, which collectively fuse to the trivial (vacuum) topological charge. Assuming the action of the symmetry does not permute anyon types, it takes the form
\begin{equation}
\label{eq:fractionalization}
\bm{R}_\mathbf{g} |\Psi_{\{a_1,\dots,a_n\}}\rangle=\prod_{j=1}^{n} \bm{U}_\mathbf{g}^{(j)} |\Psi_{\{a_1,\dots,a_n\}}\rangle
,
\end{equation}
where $\bm{U}_\mathbf{g}^{(j)}$ are unitary operators whose nontrivial action is localized in a neighborhood of the $j$th quasiparticle.
The local operators form projective representation of $G$, with multiplication given by
\begin{equation}
\bm{U}_\mathbf{g}^{(j)} \bm{U}_\mathbf{h}^{(j)}|\Psi_{\{a_1,\dots,a_n\}}\rangle=\eta_{a_j}(\mathbf{g},\mathbf{h}) \bm{U}_\mathbf{gh}^{(j)}|\Psi_{\{a_1,\dots,a_n\}}\rangle,
\end{equation}
where $\eta_a(\mathbf{g},\mathbf{h})\in {\rm U}(1)$.

The phases $\eta_a(\mathbf{g},\mathbf{h})$ must satisfy certain constraints, which provide a classification of the possible way symmetry can be fractionalized.
Since $\bm{R}_\mathbf{g} \bm{R}_\mathbf{h}=\bm{R}_\mathbf{gh}$, the fact that Eq.~(\ref{eq:fractionalization}) holds for any configuration of topological charges allowed by fusion requires that
\begin{equation}
\eta_a(\mathbf{g},\mathbf{h})\eta_b(\mathbf{g},\mathbf{h})=\eta_c(\mathbf{g},\mathbf{h}),
\end{equation}
whenever $c$ is an allowed fusion outcome of $a$ and $b$, i.e. $N_{ab}^{c} \neq 0$. This property allows us to write the projective phases as~\cite{barkeshli14}
\begin{equation}
\label{eq:cochaining_eta}
\eta_a(\mathbf{g},\mathbf{h})=M_{a,\cohosub{w}(\mathbf{g},\mathbf{h})},
\end{equation}
where $\coho{w}(\mathbf{g},\mathbf{h}) \in C^2(G,\mathcal{A})$ is an $\mathcal{A}$-valued 2-cochain, i.e. a $\mathcal{A}$-valued function on $G^2$, and
$M_{a,b}$ is the mutual braiding statistics between anyons $a$ and $b$.

Associativity of the local operators requires that
\begin{equation}
\eta_a(\mathbf{h},\mathbf{k})\eta_a(\mathbf{g},\mathbf{hk})=\eta_a(\mathbf{gh},\mathbf{k})\eta_a(\mathbf{g},\mathbf{h}).
\end{equation}
This implies $\coho{w}(\mathbf{g},\mathbf{h})\in Z^2(G,\mathcal{A})$ is a $2$-cocycle, i.e. that
\begin{equation}
\coho{w}(\mathbf{h},\mathbf{k}) \times \coho{w}(\mathbf{g},\mathbf{hk})=\coho{w}(\mathbf{gh},\mathbf{k})\times \coho{w}(\mathbf{g},\mathbf{h}).
\end{equation}

However, $\eta_a(\mathbf{g},\mathbf{h})$ have some redundancy. The local operators $\bm{U}_\mathbf{g}^{(j)}$ can be ``trivially'' redefined to $\tilde{\bm{U}}_\mathbf{g}^{(j)}$, such that
\begin{equation}
\tilde{\bm{U}}_\mathbf{g}^{(j)}|\Psi_{\{a_1,\dots,a_n\}}\rangle=\zeta_{a_j}(\mathbf{g})\bm{U}_\mathbf{g}^{(j)}|\Psi_{\{a_1,\dots,a_n\}}\rangle,
\end{equation}
where $\zeta_a(\mathbf{g})\in {\rm U}(1) $, as long as $\zeta_a(\mathbf{g})\zeta_b(\mathbf{g})=\zeta_c(\mathbf{g})$ whenever $c$ is an allowed fusion outcome of $a$ and $b$. Under this redefinition,
\begin{equation}
\tilde{\eta}_a(\mathbf{g},\mathbf{h})=\frac{\zeta_a(\mathbf{gh})}{\zeta_a(\mathbf{h})\zeta_a(\mathbf{g})}\eta_a(\mathbf{g},\mathbf{h}),
\end{equation}
and therefore projective phases $\eta_a(\mathbf{g},\mathbf{h})$ related by such transformations are physically equivalent. Since they respect fusion, the redefinition phases can similarly be written as $\zeta_a(\mathbf{g})=M_{a,\cohosub{z}(\mathbf{g})}$, where $\coho{z}(\mathbf{g})\in C^1(G,\mathcal{A})$ is a 1-cochain. In this way, the redundancy of the local operators corresponds to a redundancy of the 2-cocycles $\coho{w}(\mathbf{g},\mathbf{h})$ given by redefinition by 2-coboundaries $\text{d}\coho{z}(\mathbf{g},\mathbf{h}) = \coho{z}(\mathbf{h}) \times \overline{\coho{z}(\mathbf{g}\mathbf{h})} \times \coho{z}(\mathbf{g}) \in B^2(G,\mathcal{A})$. Thus, the possible manner in which symmetry can fractionalize, as encoded in the allowed projective phases $\eta_a(\mathbf{g},\mathbf{h})$ modulo the redundancy, is classified by the elements of the second cohomology group
\begin{equation}
[\coho{w}(\mathbf{g},\mathbf{h})]\in H^2(G,\mathcal{A})=\frac{Z^2(G,\mathcal{A})}{B^2(G,\mathcal{A})}.
\end{equation}

\subsection{$\textrm{U}(1)$ Symmetry Fractionalization}
\label{sec:u1frac}

Consider a system with an on-site $\textrm{U}(1)$ symmetry, which may be a subgroup of the full symmetry group. For example, it can be the $\textrm{U}(1)$ associated with particle number conservation or $\textrm{U}(1) < \textrm{SO}(3)$ associated with spin rotational symmetry. Let us label the elements of $\textrm{U}(1)$ as $\theta\in[0,2\pi)$, and their local action on the anyons as $\bm{U}^{(j)}_\theta$. We can choose the 2-cocycles
\begin{equation}
\label{eq:u1w}
\coho{w}(\theta_1,\theta_2)=v^{(\theta_1 + \theta_2 - [\theta_1+\theta_2]_{2\pi})/2\pi},
\end{equation}
where $v\in\mathcal{A}$, to represent the distinct cohomology classes $[\coho{w}] \in H^2(\textrm{U}(1),\mathcal{A})=\mathcal{A}$. While $\coho{w}(\theta_1,\theta_2)$ is not gauge invariant, since it can be redefined by 2-coboundaries, $v=\coho{w}(\theta,2\pi-\theta)$ is gauge invariant. Therefore, we label $\textrm{U}(1)$ fractionalization classes by $v$. Physically, the anyon $v$ is associated with the ``vison,'' which is the quasiparticle created by threading a $2\pi$ $\textrm{U}(1)$ flux~\cite{cheng15}.

Let $Q_a$ be the $\textrm{U}(1)$ charge of anyon $a$. Rotating a state by an arbitrary $\theta$ results in
\begin{equation}
    \bm{R}_\theta |\Psi_{\{a_1,\dots,a_n\}}\rangle = e^{i \theta Q}|\Psi_{\{a_1,\dots,a_n\}}\rangle,
\end{equation}
where the total charge $Q=\sum_j Q_{a_j}$ must be an integer, since $\bm{R}_{0} = \bm{R}_{2\pi}$. Meanwhile, the local operators act as
\begin{equation}
    \bm{U}_\theta^{(j)} |\Psi_{\{a_1,\dots,a_n\}}\rangle = e^{i \theta Q_{a_j}}|\Psi_{\{a_1,\dots,a_n\}}\rangle,
\end{equation}
where $Q_a$ need not be integers. This action of $\bm{U}_\theta^{(j)}$ is not gauge invariant, but a gauge invariant statement can be obtained by applying a complete 2$\pi$ rotation [with the use of Eq.~(\ref{eq:cochaining_eta})]:
\begin{equation}
 \bm{U}_\theta^{(j)}\bm{U}_{2\pi-\theta}^{(j)} |\Psi_{\{a_1,\dots,a_n\}}\rangle = M_{a_j, v} |\Psi_{\{a_1,\dots,a_n\}}\rangle.
\end{equation}
Thus, the anyon $a$ has a possibly fractional charge $Q_a$, which is given by the relation
\begin{equation}
\label{eq:fraccharge}
 e^{i 2\pi Q_a}=M_{a,v}.
\end{equation}

\subsection{Translational Symmetry Fractionalization}

Consider a 2D system in a topological phase with $\mathbb{Z}^2$ translational symmetry. The fractionalization of this symmetry requires a straightforward modification of the on-site formalism. In particular, the state vector $|\Psi_{\{a_1,\dots,a_n\}}\rangle$ on the right hand side of Eq.~(\ref{eq:fractionalization}) must have the positions of its quasiparticles translated (according to the applied translation operator) with respect to $|\Psi_{\{a_1,\dots,a_n\}}\rangle$ on the left hand side, and the local unitary operators $\bm{U}^{(j)}_{\bf g}$ should be understood to act nontrivially in a neighborhood of the translated quasiparticle positions~\cite{barkeshli14}. Let us label the generators of translation as $T_x$ and $T_y$ and their corresponding local unitary operators as $\bm{U}^{(j)}_x$ and $\bm{U}^{(j)}_y$.

We can choose the 2-cocycles
\begin{equation}
    \coho{w}(T_x^{m_x}T_y^{m_y},T_x^{n_x}T_y^{n_y})=b^{m_y n_x},
\end{equation}
where $b\in\mathcal{A}$, to represent the distinct cohomology classes $[\coho{w}] \in H^2(\mathbb{Z}^{2},\mathcal{A})=\mathcal{A}$. While $\coho{w}(T_x^{m_x}T_y^{m_y},T_x^{n_x}T_y^{n_y})$ is not gauge invariant, since it can be redefined by 2-coboundaries, the quantity $\coho{w}(T_y,T_x)\times\overline{\coho{w}(T_x,T_y)} = b$ is gauge invariant and, moreover, completely specifies cohomology class. Therefore, we can label translational symmetry fractionalization classes by $b\in\mathcal{A}$.

Physically, the anyon $b$ can be thought of as the background anyonic flux per unit cell. This is because
\begin{align}
    &(\bm{U}^{(j)}_{T_y})^{-1}(\bm{U}^{(j)}_{T_x})^{-1} \bm{U}^{(j)}_{T_y} \bm{U}^{(j)}_{T_x}  |\Psi_{\{a_1,\dots,a_n\}}\rangle \nonumber \\
    & \quad =\frac{\eta_{a_j}(T_y,T_x)}{\eta_{a_j}(T_x,T_y)} |\Psi_{\{a_1,\dots,a_n\}}\rangle \nonumber \\
    & \quad =M_{a_j,\cohosub{w}(T_y,T_x)} M_{a_j,\overline{\cohosub{w}(T_x,T_y)}}|\Psi_{\{a_1,\dots,a_n\}}\rangle \nonumber \\
    & \quad =M_{a_j,b}|\Psi_{\{a_1,\dots,a_n\}}\rangle.
\end{align}
That is, when an anyon $a$ is transported around a unit cell, the wavefunction acquires a phase corresponding to braiding $a$ around $b$.

\subsection{Flux Threading Argument}
\label{sec:fluxthreading}

\begin{figure}
    \centering
    \includegraphics{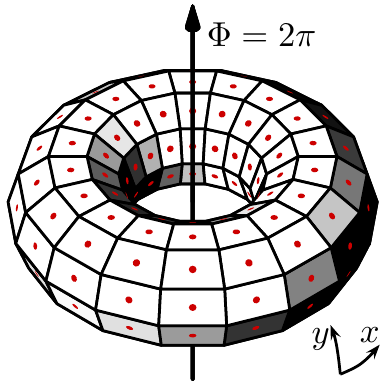}
    \includegraphics[trim= -10 -6 0 0]{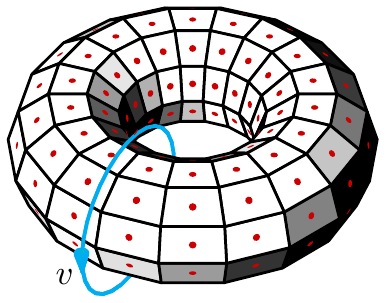}
    \\$\bm{\mathcal{G}}|\Psi(2\pi)\rangle=\bm{W}_v|\Psi(0)\rangle$
    \caption{Threading a $2\pi$ flux through the handle of the torus creates a $v$ anyon loop (blue). The dots represent the anyonic flux per unit cell $b$ (red).}\label{fig:fractorus}
    \includegraphics[trim= 0 0 0 -5]{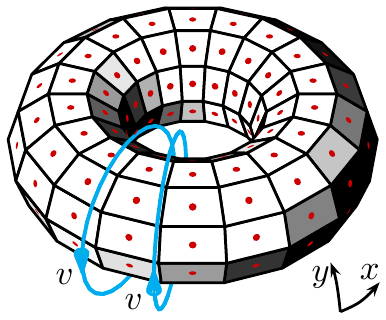}
    \includegraphics[trim= -10 -9 0 -5]{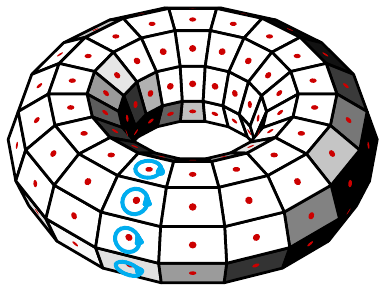}
    \\$(\bm{R}_{T_x})^{-1}(\bm{W}_v)^{-1}\bm{R}_{T_x}\bm{W}_v|\Psi(0)\rangle=(M_{v, b})^{L_y}|\Psi(0)\rangle$
    \caption{The $\bm{R}_{T_x}$ eigenvalue of $\bm{W}_v|\Psi(0)\rangle$ is determined by the mutual braiding statistics between $v$ and $b$. To go from the l.h.s. to the r.h.s., we partially fused the adjacent $v$ anyon loops, being careful not to pass them through the anyonic flux $b$ lines emanating from the center of every cell of the torus.}\label{fig:fractorus2}
\end{figure}

Consider a system with both on-site $\textrm{U}(1)$ symmetry and $\mathbb{Z}^2$ translational symmetry. Using K\"unneth formula for group cohomology, one has~\cite{cheng15}
\begin{equation}
    H^2(\textrm{U}(1)\times \mathbb{Z}^2,\mathcal{A})=H^2(\textrm{U}(1),\mathcal{A})\times H^2(\mathbb{Z}^2,\mathcal{A}),
\end{equation}
which means that the fractionalization for the combined symmetries are determined by that of the $\textrm{U}(1)$ and translational symmetries, which can be independently specified. Suppose the system belongs to $\textrm{U}(1)$ fractionalization class $v$ and translational symmetry fractionalization class $b$.

If we consider a state that has $2\pi$ $\textrm{U}(1)$ flux through a handle of the torus and transport an anyon $a$ around the handle, so that it winds around the flux once, the wavefunction will acquire the Aharanov-Bohm phase $e^{i 2\pi Q_a}$. By Eq.~(\ref{eq:fraccharge}), this is identical to the phase $M_{a,v}$ that is acquired by braiding $a$ around $v$. Therefore, the effect of threading the flux through a handle of the torus should be gauge equivalent to creating a vison loop that wraps around the handle, as illustrated in Fig.~\ref{fig:fractorus}. That is
\begin{equation}
\bm{\mathcal{G}}|\Psi(2\pi)\rangle=\bm{W}_v|\Psi(0)\rangle,
\end{equation}
where $\bm{W}_v$ is an operator that creates a $v$ anyon loop wrapping around the handle of the torus.~\footnote{The application of $\bm{W}_v$ to a ground state is equivalent to creating a $v-\bar{v}$ pair of anyons, adiabatically transporting $v$ around the cycle of the torus, and then annihilating the pair to vacuum.} (See Appendix~\ref{ap:fluxthreading} for a more direct argument.)

The state $\bm{W}_v |\Psi(0)\rangle$ has momentum $P_x(0)+ 2\pi Q_{b} L_y \mod 2\pi$, since
\begin{equation}
    (\bm{R}_{T_x})^{-1}(\bm{W}_v)^{-1} \bm{R}_{T_x} \bm{W}_v|\Psi(0)\rangle= (M_{v,b})^{L_y}|\Psi(0)\rangle,
\end{equation}
which can be understood from the relation in Fig.~\ref{fig:fractorus2}. On the other hand, we know the state $\bm{\mathcal{G}}|\Psi(2\pi)\rangle$ has momentum $P_x(0)+2\pi\nu L_y \mod 2\pi$. Equating the momenta of $\bm{\mathcal{G}}|\Psi(2\pi)\rangle$ and $\bm{W}_v |\Psi(0)\rangle$, and repeating the argument in the other direction yields
\begin{equation}
\label{eq:topoluttinger}
    \nu=Q_{b}\equiv\nu_\text{topo} \mod 1.
\end{equation}
In other words, the filling fraction of a 2D SET phase is equal to the $\textrm{U}(1)$ charge of the background anyonic flux per unit cell.

Eq.~(\ref{eq:topoluttinger}), which relates microscopic and emergent properties of the system, can be viewed as a constraint on the allowed SET order that may exist at a given filling. For example, consider the Ising anyon model, which contains Abelian anyons $I$ and $\psi$, and non-Abelian anyon $\sigma$. The fact that $M_{I,I}=M_{I,\psi}=M_{\psi,\psi}=1$ implies that $Q_{b}=0$ for any fractionalization pattern, and so it is impossible to have the pure Ising topological order at a non-integer fractional filling.

\section{Fractionalized Fermi Liquid}
\label{sec:ffl}

A fractionalized Fermi liquid is a gapless system with $\textrm{U}(1)$ and translational SET order, whose gapless modes are well-described by Fermi liquid theory, and whose symmetries are fractionalized. We assume that topological excitations and gapless excitations coexist, but are effectively decoupled from one another, i.e. the system decouples into an SET sector and a Fermi liquid sector, and is consequently in a strong quasi-topological phase~\cite{bonderson13}. We consider a 2D system for which the SET order belongs to $\textrm{U}(1)$ fractionalization class $v$ and translational symmetry fractionalization class $b$.

Similar to the situation described in Sec.~\ref{sec:fluxthreading}, starting from a ground state $|\Psi(0)\rangle$ of a fractionalized Fermi liquid and threading a $2\pi$ flux through the handle of a torus is gauge equivalent to applying a vison loop that wraps around the handle to the state $|\Psi'(0)\rangle$,
\begin{equation}
    \bm{\mathcal{G}}|\Psi(2\pi)\rangle=\bm{W}_v|\Psi'(0)\rangle,
\end{equation}
where $|\Psi'(0)\rangle$ is $|\Psi(0)\rangle$ with a Galilean boosted Fermi sea, so that it is in the same topological sector as $|\Psi(0)\rangle$, but has a shifted momentum. Note that the assumption of the decoupling between the SET sector and the Fermi liquid sector is crucial here, since it allows us to separate the effect of flux threading on the SET sector, i.e. wrapping a vison loop around the handle, from its effect on the Fermi liquid sector, i.e. boosting the Fermi sea. If the topological excitations were to interact with the Fermi liquid quasiparticles in a manner that nontrivially coupled the SET sector and the Fermi liquid sector, then the effect of flux threading may not be so cleanly separable.

The state $|\Psi'(0)\rangle$ has momentum $P_x(0)+2\pi N_F/L_x \mod 2\pi$, due to the Fermi liquid quasiparticles. As explained in Sec.~\ref{sec:fluxthreading}, the state $\bm{W}_v|\Psi'(0)\rangle$ has momentum $2\pi Q_{b}L_y$ relative to the state $|\Psi'(0)\rangle$. On the other hand, we know the state $\bm{\mathcal{G}}|\Psi(2\pi)\rangle$ has momentum $P_x(0)+2\pi\nu L_y \mod 2\pi$. Equating the momenta of $\bm{\mathcal{G}}|\Psi(2\pi)\rangle$ and $\bm{W}_v|\Psi'(0)\rangle$ and repeating the argument in the other direction yields Luttinger's theorem for a 2D fractionalized Fermi liquid:
\begin{equation}
    \nu=Q_{b}+\frac{V_F}{(2\pi)^2} \mod 1.
\end{equation}
This is essentially a combination of Eq.~(\ref{eq:luttinger}) and Eq.~(\ref{eq:topoluttinger}). We see that the background anyonic flux can appropriate some of the charge available to the emergent degrees of freedom, thus changing the Fermi volume. Or, put differently, the Fermi volume is determined by the filling fraction of the Fermi liquid sector:
\begin{equation} \label{eq:fflluttinger}
    \nu-\nu_\text{topo}=\frac{V_F}{(2\pi)^2} \mod 1.
\end{equation}

\subsection{$\mathbb{Z}_2$ Fractionalized Fermi Liquid: FL*}

Consider a 2D periodic lattice with $\nu_c=\nu_{c\uparrow}+\nu_{c\downarrow}$ conduction electrons and $\nu_s$ spin-$\frac{1}{2}$ localized spins per unit cell, governed by the Kondo model Hamiltonian
\begin{align}
    \bm{H}&=-t\sum_{\langle jk \rangle,\alpha}(\bm{c}_{j\alpha}^\dagger \bm{c}_{k\alpha}+\text{h.c.})+U\sum_j \bm{n}_{j\uparrow} \bm{n}_{j\downarrow} \notag \\
    &\qquad +K\sum_j \vec{\bm{s}}_j\cdot\vec{\bm{S}}_j +J \sum_{\langle jk \rangle} \vec{\bm{S}}_j\cdot\vec{\bm{S}}_k,
\end{align}
where $\vec{\bm{s}}_j=\sum_{\alpha\beta}\bm{c}^\dagger_{j\alpha}\vec{\sigma}_{\alpha\beta}\bm{c}_{j\beta}/2$.
As explained in App.~\ref{ap:kondo}, the above Hamiltonian has two global $\textrm{U}(1)$ symmetries, denoted $\textrm{U}(1)_\uparrow$ and $\textrm{U}(1)_\downarrow$, which correspond to the independently conserved quantities $\nu_{c\uparrow}+m_s \nu_s$ and $\nu_{c\downarrow}-m_s\nu_s$, respectively, where $m_s$ is the magnetization per localized spin.

In the ordinary Fermi liquid phase of the Kondo model, these two symmetries lead to the Luttinger's theorems
\begin{align}
\label{eq:kondoluttinger2}
\nu_{c\uparrow}+\left(\frac{1}{2}+m_s\right)\nu_s &= \frac{ V_{F\uparrow}}{(2\pi)^2} \mod 1,\\
\label{eq:kondoluttinger3}
\nu_{c\downarrow}+\left(\frac{1}{2}-m_s\right)\nu_s &= \frac{ V_{F\downarrow}}{(2\pi)^2} \mod 1,
\end{align}
which can be combined to give the spin-summed Luttinger's theorem
\begin{equation}
\label{eq:kondoluttinger}
    \nu_c+\nu_s=\frac{V_F}{(2\pi)^2} \mod 2.
\end{equation}
See App.~\ref{ap:kondo} for details.

If the Kondo model is placed on a geometrically frustrated lattice, e.g. triangular lattice, then at low temperatures and small enough values of $K$, the localized spins are believed to topologically order. In this case, the system may enter the so-called FL* phase of the Kondo model~\cite{senthil03, chowdhury14, thomson15}, which is a fractionalized Fermi liquid. Let us assume that $K=0$, so that the electrons are decoupled from the spins, and that the spins form a $\mathbb{Z}_2$ spin liquid with toric code topological order.

In this case, the localized spins carry $\textrm{U}(1)_{\uparrow}$ and $\textrm{U}(1)_{\downarrow}$ charge values of 1/2. Consequently, the Luttinger's theorems are modified by $\nu_{\text{topo}}=\nu_s/2$ to give
\begin{align}
    \nu_{c\uparrow}+m_s\nu_s &= \frac{ V_{F\uparrow}}{(2\pi)^2} \mod 1,\\
    \nu_{c\downarrow}-m_s\nu_s&= \frac{ V_{F\downarrow}}{(2\pi)^2} \mod 1,
\end{align}
which can be combined to give the spin-summed Luttinger's theorem for the FL* phase
\begin{equation}
\label{eq:kondoluttingerfl*}
    \nu_c=\frac{V_F}{(2\pi)^2} \mod 2 .
\end{equation}
We emphasize that this result for the FL* phase differs from the result in Eq.~(\ref{eq:kondoluttinger}) for the ordinary Fermi liquid phase when the number of localized spins per unit cell $\nu_s$ is odd. This difference can be understood by studying symmetry fractionization of the $\mathbb{Z}_2$ spin liquid, as we now explain in more detail.

Recall that the toric code~\cite{kitaev03} has four types of anyons: trivial excitations $I$, bosons $e$ and $m$, and fermionic composites $f=e\times m$. They are all Abelian and obey $\mathbb{Z}_2\times\mathbb{Z}_2$ fusion rules. The nontrivial braiding statistics are $M_{e,m}=M_{e,f}=M_{m,f}=-1$. Let the $\textrm{U}(1)_\uparrow$ and $\textrm{U}(1)_\downarrow$ symmetry fractionalization class be specified by $v=m$. In this case, $Q_{I}=Q_{m}=0$ and $Q_{e}=Q_{f}=1/2$, i.e. $e$ is a spin-$\frac{1}{2}$ spinon, $m$ is a spinless vison, and $f$ is a spin-$\frac{1}{2}$ fermion.~\footnote{Observe that the spin symmetry of the localized spins has been fractionalized. To see this, first note that while the localized spins have $\textrm{SU}(2)$ spin symmetry, the relevant symmetry group is actually $\textrm{SU}(2)$ modded by its center $\mathbb{Z}_2$: $\textrm{SU}(2)/\mathbb{Z}_2=\textrm{SO}(3)$. In other words, the Hilbert space factorizes into two disjoint subspaces, and the $\textrm{SU}(2)$ operators act as a direct product of $\textrm{SO}(3)$ operators on each subspace. The representations of $\textrm{SO}(3)$ are classified by $\mathbb{Z}$, i.e. integer spins. Indeed, local operators such as $\bm{S}^+_j$ and $\bm{S}^-_j$ change the spin by an integer. Thus, an excitation must normally have integer spin. However, in the $\mathbb{Z}_2$ spin liquid, the $e$ and $f$ are spin-$\frac{1}{2}$ excitations.

Alternatively, we may view the localized spins as electrons, in which case the relevant symmetry group is the product of the $\textrm{SU}(2)$ spin symmetry group and the $\textrm{U}(1)$ charge symmetry group, modded by its center $\mathbb{Z}_2$: $\textrm{SU}(2)\times $\textrm{U}(1)$/\mathbb{Z}_2=\textrm{U}(2)$. The representations of $\textrm{U}(2)$ are classified by $(s,q)$, where $2s$ and $q$ are integers whose sum is even. Therefore, a particle with spin $s$ and charge $q$ must obey $2s+q \mod 2=0$. The electron, for instance, transforms as the $(\frac{1}{2}, 1)$ representation. However, in the $\mathbb{Z}_2$ spin liquid, the $e$ and $f$ are spin-$\frac{1}{2}$ chargeless excitations.}
If $\nu_s$ is even, then, by Eq. (\ref{eq:topoluttinger}), the translational symmetry fractionalization class is either $b=I$ or $b=m$, and Eq.~(\ref{eq:kondoluttinger}) agrees with Eq.~(\ref{eq:kondoluttingerfl*}). However, if $\nu_s$ is odd, then the translational symmetry fractionalization class is either $b=e$ or $b=f$. In this case, Eq.~(\ref{eq:kondoluttinger}) and Eq.~(\ref{eq:kondoluttingerfl*}) clearly disagree, and the topological enrichment of Luttinger's theorem for the FL* phase is manifest.

\subsection{3D $\mathbb{Z}_2$ Fractionalized Fermi Liquid}

\begin{figure}
    \centering
    \includegraphics{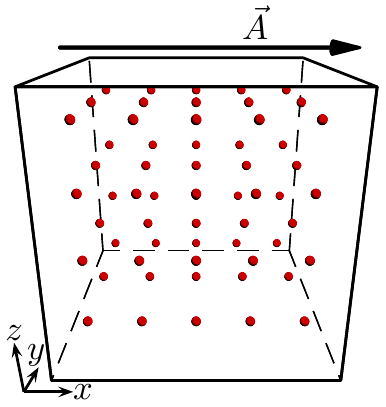}
    \includegraphics[trim= -10 -6 0 0]{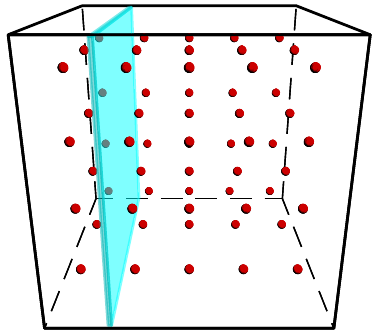}
    \\$\bm{\mathcal{G}}|\Psi(2\pi)\rangle=\bm{W}_m|\Psi'(0)\rangle$
    \caption{Inserting a $2\pi$ flux along the $x$ direction is equivalent to having an $m$ membrane (blue) in the $yz$ plane. The dots represent an $e$ occupying every cell (red). (The underlying 3D periodic lattice is not shown.)}\label{fig:fractorus3d}
    \includegraphics[trim=0 0 0 -5]{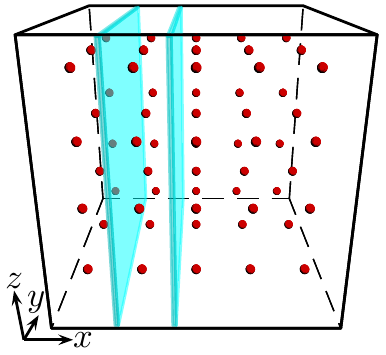}
    \includegraphics[trim= -10 -6 0 -5]{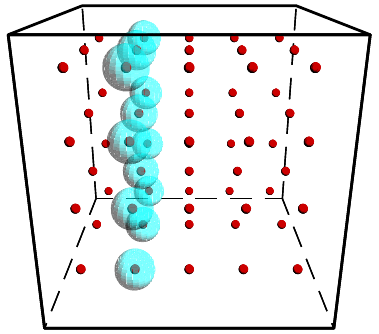}
    \\$(\bm{R}_{T_x})^{-1}(\bm{W}_m)^{-1}\bm{R}_{T_x}\bm{W}_m|\Psi(0)\rangle=(M_{m, e})^{L_y L_z}|\Psi(0)\rangle$
    \caption{The $\bm{R}_{T_x}$ eigenvalue of $\bm{W}_m|\Psi(0)\rangle$ is determined by the mutual braiding statistics between $m$ and $e$. To go from the l.h.s. to the r.h.s., we have partially fused the adjacent $m$ membranes, being careful not to pass them through the $e$ at the center of every cell.}\label{fig:fractorus3d2}
\end{figure}

Consider a spinless Fermi liquid that is accompanied by 3D bosonic toric code ($\mathbb{Z}_{2}$ gauge theory) topological order. The 3D bosonic toric code has four types of topological excitations: trivial excitations $I$, point excitations $e$, loop excitations $m$, and composite loop excitations $f=e\times m$~\cite{hamma05, keyserlingk13}. Topological loop excitations are created along the boundary of a fluctuating surface operator, similar to how topological point excitations are created at the endpoints of fluctuating string operators. While the fusion rules and exchange statistics of excitations in a 3D topological order are more intricate than in 2D, our discussion will simply rely on the fact that $M_{e,m}=-1$, which now expresses the phase obtained by taking an $e$ quasiparticle around a circuit that links a $m$ loop once. As in 2D, the topological enrichment of Luttinger's theorem can be understood in terms of the symmetry fractionalization. Although 3D symmetry fractionalization currently lacks a general formalism, it has been recently studied for the 3D toric code~\cite{cheng15b}.

Suppose that $\textrm{U}(1)$ charge symmetry is fractionalized such that $e$ is charge-$\frac{1}{2}$, and translational symmetry is fractionalized such that an $e$ occupies every unit cell, and hence we have $\nu_{\text{topo}}=1/2$. Then, at least for our purposes, inserting a flux such that an uniform electric field is induced along the $x$-direction is gauge equivalent to introducing an $m$ membrane in the $yz$-plane, as shown in Fig.~\ref{fig:fractorus3d}, since transporting an $e$ anyon along the $x$-direction around a nontrivial cycle of the torus results in the wavefunction acquiring a phase of $-1$ in both cases. (Since the $m$ membrane is created by a non-contractible surface operator $\bm{W}_m$ that has no boundary, it does not create an $m$ loop excitation, but still acts nontrivially on the ground states.) Mathematically, this is expressed as $\bm{\mathcal{G}}|\Psi(2\pi)\rangle=\bm{W}_m|\Psi'(0)\rangle$, where $|\Psi'(0)\rangle$ is in the same topological sector as $|\Psi(0)\rangle$.

The state $|\Psi'(0)\rangle$ has momentum $P_x(0)+2\pi N_F/L_x \mod 2\pi$, due to the Fermi liquid quasiparticles. The state $\bm{W}_m|\Psi'(0)\rangle$ has momentum $\pi L_y L_z$ relative to $|\Psi'(0)\rangle$, since
\begin{equation}
    (\bm{R}_{T_x})^{-1}(\bm{W}_m)^{-1} \bm{R}_{T_x} \bm{W}_m|\Psi(0)\rangle= (M_{m,e})^{L_y L_z}|\Psi(0)\rangle,
\end{equation}
which can be understood from the relation in Fig.~\ref{fig:fractorus3d2}. On the other hand, we know that the state $\bm{\mathcal{G}}|\Psi(2\pi)\rangle$ has momentum $P_x(0)+2\pi\nu L_y L_z \mod 2\pi$. Equating the momenta of $\bm{\mathcal{G}}|\Psi(2\pi)\rangle$ and $\bm{W}_v |\Psi(0)\rangle$ and repeating the argument in the other directions yields the topologically enriched Luttinger's theorem
\begin{equation}
    \nu-\frac{1}{2}=\frac{V_F}{(2\pi)^3} \mod 1.
\end{equation}

In the context of the 3D Kondo model, a fractionalized Fermi liquid phase is realized when the localized spins acquire the 3D bosonic toric code topological order. If the $\textrm{U}(1)_\uparrow$ and $\textrm{U}(1)_\downarrow$ symmetries are fractionalized such that $e$ quasiparticles carry spin-$\frac{1}{2}$, and the translational symmetry is fractionalized such that an $e$ (or $f$) occupies each unit cell, then the topologically enriched Luttinger's theorem is
\begin{equation}
\nu_c+\nu_s-1=\frac{V_F}{(2\pi)^3} \mod 2.
\end{equation}

\section{Discussion}
\label{sec:discussion}

We have extended Oshikawa's arguments to systems that possess SET order. For fractionalized Fermi liquids, this led to a topologically enriched version of Luttinger's theorem. The modified Luttinger's theorem of Eq.~(\ref{eq:TELutt}) determines how the presence of topological order can change the Fermi volume. From the opposite perspective, this relation places strict constraints on the possible SET order allowed to manifest in a fractionalized Fermi liquid with an experimentally observed Fermi volume that deviates from the na\"ive value expected for an ordinary Fermi liquid.

While we have focused on systems whose SET sector and Fermi liquid sector are effectively decoupled, it would be interesting to apply our arguments to other gapless topological systems, e.g. $\mathbb{Z}_2$ and $\textrm{U}(1)$ gapless spin liquids. For gapless spin liquids, the challenge is understanding their symmetry fractionalization and their behavior under flux threading, particularly when there are nontrivial interactions between the gapless topological excitations and the gapless Fermi liquid quasiparticles. In general, it would be interesting to relax our assumption that the SET sector and Fermi liquid sector of a fractionalized Fermi liquid are decoupled. Introducing some interaction that mixes these sectors would drive the system into a weak quasi-topological phase, and may nontrivially modify our results.

Finally, a natural extension of our arguments would be to fully understand their generalization to higher dimensional systems. As mentioned, we expect Eq.~(\ref{eq:TELutt}) to hold for a general $D$-dimensional fractionalized Fermi liquid, but our ability to establish this relation is limited by the fact that the theory of higher dimensional topological order and symmetry enrichment is not yet fully developed.

\begin{acknowledgments}
We thank C. Nayak and M. M. Zaletel for useful discussions and comments on this work.
\end{acknowledgments}

\appendix

\section{Luttinger's Theorem for Kondo Model}
\label{ap:kondo}

Consider a 2D periodic lattice with $\nu_c=\nu_{c\uparrow}+\nu_{c\downarrow}$ conduction electrons and $\nu_s$ spin-$S$ localized spins per unit cell, governed by the translationally invariant Kondo model Hamiltonian
\begin{widetext}
\begin{eqnarray}
\bm{H}&=& -t\sum_{\langle jk \rangle,\alpha}(\bm{c}_{j\alpha}^\dagger \bm{c}_{k\alpha}+\text{h.c.})+U\sum_j \bm{n}_{j\uparrow} \bm{n}_{j\downarrow}
+K\sum_j \vec{\bm{s}}_j\cdot\vec{\bm{S}}_j +J \sum_{\langle jk \rangle} \vec{\bm{S}}_j\cdot\vec{\bm{S}}_k  \\
&=& -t\sum_{\langle jk \rangle,\alpha}(\bm{c}_{j\alpha}^\dagger \bm{c}_{k\alpha}+\text{h.c.})+U\sum_j \bm{n}_{j\uparrow} \bm{n}_{j\downarrow}
+\frac{1}{2}K\sum_j \left[ (\bm{n}_{j\uparrow}-\bm{n}_{j\downarrow})\bm{S}^z_j+\bm{c}^\dagger_{j\downarrow} \bm{c}_{j\uparrow} \bm{S}^+_j +\bm{c}^\dagger_{j\uparrow} \bm{c}_{j\downarrow} \bm{S}^-_j \right] \notag \\
&& \qquad \qquad +J \sum_{\langle jk \rangle} \left[ \bm{S}^z_j \bm{S}^z_k + \frac{1}{2}(\bm{S}^+_j\bm{S}^-_k + \text{h.c.}) \right],
\end{eqnarray}
where $\vec{\bm{s}}_j=\sum_{\alpha\beta}\bm{c}^\dagger_{j\alpha}\vec{\sigma}_{\alpha\beta}\bm{c}_{j\beta}/2$.

The above Hamiltonian has two global $\textrm{U}(1)$ symmetries, corresponding to the conserved quantities $\nu_{c\uparrow}+m_s\nu_s$ and $\nu_{c\downarrow}-m_s\nu_s$, where $m_s$ is the magnetization per localized spin. The first of these, which we denote as
$\textrm{U}(1)_{\uparrow}$, is generated by the transformations: $\bm{c}^\dagger_{j\uparrow}\rightarrow e^{i \theta}\bm{c}^\dagger_{j\uparrow}$, $\bm{c}_{j\uparrow}\rightarrow e^{-i \theta}\bm{c}_{j\uparrow}$, and $\bm{S}^\pm_j\rightarrow e^{\pm i \theta}\bm{S}^\pm_j$. This global symmetry can be promoted to a local symmetry by introducing a gauge field $A_{jk}$ that couples to spin up electrons and the localized spins, modifying the Hamiltonian to
\begin{eqnarray}
\bm{H}'&=& -t\sum_{\langle jk \rangle}(e^{i A_{jk}}\bm{c}_{j\uparrow}^\dagger \bm{c}_{k\uparrow}+\bm{c}_{j\downarrow}^\dagger \bm{c}_{k\downarrow}+\text{h.c.})+U\sum_j \bm{n}_{j\uparrow} \bm{n}_{j\downarrow}
+\frac{1}{2}K\sum_j \left[ (\bm{n}_{j\uparrow}-\bm{n}_{j\downarrow})\bm{S}^z_j+\bm{c}^\dagger_{j\downarrow} \bm{c}_{j\uparrow} \bm{S}^+_j +\bm{c}^\dagger_{j\uparrow} \bm{c}_{j\downarrow} \bm{S}^-_j \right] \notag \\
&& \qquad \qquad +J \sum_{\langle jk \rangle} \left[ \bm{S}^z_j \bm{S}^z_k + \frac{1}{2}(e^{iA_{jk}}\bm{S}^+_j\bm{S}^-_k + \text{h.c.}) \right],
\end{eqnarray}
\end{widetext}
which now has the local $\textrm{U}(1)_{\uparrow}$ symmetry given by the transformations: $\bm{c}_{j\uparrow}\rightarrow e^{i \theta_j}\bm{c}_{j\uparrow}$, $\bm{c}_{j\uparrow}^\dagger\rightarrow e^{-i \theta_j}\bm{c}_{j\uparrow}^\dagger$, $\bm{S}^\pm_j\rightarrow e^{\pm i \theta_j}\bm{S}^\pm_j$, and $A_{jk}\rightarrow A_{jk}+\theta_k-\theta_j$.~\footnote{This $\textrm{U}(1)$ symmetry may seem artificial. A more physical viewpoint is to let the conduction electrons have a $\textrm{U}(1)_{S^z}\times \textrm{U}(1)_c$ spin and charge symmetry, and the localized spins have a $\textrm{U}(1)_{S^z}$ spin symmetry. In this case, we carry out Oshikawa's argument by threading $2\pi$ $\textrm{U}(1)_{S^z}$ flux and $\pi$ $\textrm{U}(1)_c$ flux, so that the spin up electrons experience a $2\pi(+\frac{1}{2})+\pi=2\pi$ flux, the spin down electrons experience a $2\pi(-\frac{1}{2})+\pi=0$ flux, and the localized spins experience a $2\pi S$ flux.}

Consider starting in a ground state $|\Psi(0)\rangle$ with $\bm{R}_{T_x}$ eigenvalue $e^{i P_x(0)}$ and threading a $2\pi$ $\textrm{U}(1)_{\uparrow}$ flux through the handle of the torus. This can be accomplished by tuning the vector potential from $\vec{A}(0)=(0,0)$ to $\vec{A}(2 \pi)=(2\pi/ L_x,0)$, i.e. $A_{jk}=\int_{\vec{r}_j}^{\vec{r}_k} d\vec{r}\cdot \vec{A}=\frac{[\vec{r}_j-\vec{r}_k]_x}{L_x}$ after the flux insertion. Although $\bm{H}'(2\pi)\ne\bm{H}'(0)$, the large gauge transformation
\begin{equation}
\bm{\mathcal{G}}_\uparrow= e^{ i 2\pi \sum_{j} \frac{[\vec{r}_j]_x}{L_x}(\bm{n}_{j\uparrow}+\bm{S}^z_j) },
\end{equation}
removes the flux, i.e. $\bm{\mathcal{G}}_\uparrow\bm{H}'(2\pi)\bm{\mathcal{G}}_\uparrow^{-1}=\bm{H}'(0)$. Therefore, the state $\bm{\mathcal{G}}_\uparrow|\Psi(2\pi)\rangle$ must be an eigenstate of $\bm{H}'(0)$.

Since $[\bm{R}_{T_x},\bm{H}'(\Phi)]=0$ throughout the flux threading process, $|\Psi(2\pi)\rangle$ has momentum $P_x(0)$, and since
\begin{equation}
    \bm{\mathcal{G}}_\uparrow^{-1} \bm{R}_{T_x} \bm{\mathcal{G}}_\uparrow = \bm{R}_{T_x} e^{i 2\pi \left[ \frac{1}{L_x}\sum_j (\bm{n}_{j\uparrow}+\bm{S}^z_j)+\sum_{j|[\vec{r}_j]_x=1} \bm{S}^z_j \right]},
\end{equation}
the state $\bm{\mathcal{G}}_\uparrow|\Psi(2\pi)\rangle$ has momentum $P_x(0)+2\pi \left[ \nu_{c\uparrow}+(S+m_s)\nu_s \right] L_y \mod 2\pi$.

This shift can be compared with the momentum shift of the emergent degrees of freedom. Assuming the system is a spinful Fermi liquid, threading the flux shifts the momentum of each of the $N_{F\uparrow}$ spin up quasiparticles by $2\pi/L_x$.

Equating the two momentum shifts and repeating the argument in the other direction yields Luttinger's theorem for spin up quasiparticles:
\begin{equation}
    \nu_{c\uparrow}+ (S+m_s)\nu_s=\frac{V_{F\uparrow}}{(2\pi)^2} \mod 1
\end{equation}
where $V_{F\uparrow}\equiv(2\pi)^2 N_{F\uparrow}/L_x L_y$ is the Fermi volume.

The $\textrm{U}(1)_{\downarrow}$ symmetry is defined similarly for the spin down quasiparticles, and the same arguments give the corresponding Luttinger's theorem:
\begin{equation}
    \nu_{c\downarrow}+ (S-m_s)\nu_s= \frac{ V_{F\downarrow}}{(2\pi)^2} \mod 1.
\end{equation}
Combining these results and using the fact that the number of filled bands for spin up electrons is equal to the number for spin down electrons gives the spin-summed Luttinger's theorem:
\begin{equation}
    \nu_c+2S\nu_s=\frac{V_F}{(2\pi)^2} \mod 2.
\end{equation}

\section{Gauge Equivalence Between Flux Threading and Creating Anyon Loop}
\label{ap:fluxthreading}

In this appendix, we establish the equivalence between adiabatically threading a $2\pi$ $\textrm{U}(1)$ through a handle of the torus and creating a vison $v$ anyonic flux loop around the handle. Consider a system with on-site $\textrm{U}(1)$ symmetry. Starting with the Hamiltonian $\bm{H}(0)$ and state $|\Psi(0)\rangle$, threading a $2\pi$ $\textrm{U}(1)$ flux through the handle of the torus results in the Hamiltonian $\bm{H}(2\pi)$ and state $|\Psi(2\pi)\rangle$, where $\bm{H}(2\pi)=\bm{\mathcal{G}}^{-1}\bm{H}(0)\bm{\mathcal{G}}$ for some large gauge transformation
\begin{equation}
    \bm{\mathcal{G}}=e^{i2\pi\sum_j \frac{[\vec{r}_j]_x}{L_x} \bm{q}_j}=\prod_j \bm{R}^{(j)}_{2\pi [\vec{r}_j]_x/L_x}.
\end{equation}
Here, $\bm{q}_j$ measures the $\textrm{U}(1)$ charge of site $j$, while $\bm{R}^{(j)}_\theta=e^{i\theta \bm{q}_j}$ rotates it by $\theta$. The state of the system after threading the flux and applying the gauge transformation is $\bm{\mathcal{G}}|\Psi(2\pi)\rangle$.

For simplicity, let us assume that the Hamiltonian consists of on-site and nearest neighbor terms only, i.e.
\begin{equation}
    \bm{H}(0)=\sum_j \bm{h}_j + \sum_{\langle jk\rangle} \bm{h}_{jk},
\end{equation}
and investigate its transformation under $\bm{\mathcal{G}}^{-1}\bm{H}(0)\bm{\mathcal{G}}$.
Since
\begin{equation}
\bm{R}^{(j)}_{-\theta} \bm{h}_j \bm{R}^{(j)}_\theta = \bm{R}_{-\theta} \bm{h}_j \bm{R}_\theta = \bm{h}_j,
\end{equation}
the on-site terms in the Hamiltonian are unaffected by $\bm{\mathcal{G}}$:
\begin{equation}\label{eq:on-sitetrans}
     \bm{\mathcal{G}}^{-1} \bm{h}_j \bm{\mathcal{G}}=\bm{h}_j.
\end{equation}
Similarly, since
\begin{equation}
\bm{R}^{(j)}_{-\theta}\bm{R}^{(k)}_{-\theta} \bm{h}_{jk}\bm{R}^{(j)}_\theta\bm{R}^{(k)}_\theta=\bm{h}_{jk},
\end{equation}
the $y$-direction nearest neighbor terms in the Hamiltonian are unaffected. Only the $x$-direction nearest neighbor terms are transformed nontrivially by $\bm{\mathcal{G}}$. Specifically, we have
\begin{align}
\label{eq:nntrans}
    \bm{\mathcal{G}}^{-1}\bm{h}_{jk}\bm{\mathcal{G}}=\left\{
    \begin{array}{rl}
        \bm{R}^{(k)}_{2\pi/L_x} \bm{h}_{jk} \bm{R}^{(k)}_{-2\pi/L_x} & \text{if $\vec{r}_j-\vec{r}_k=(1,0)$,}\\
        \bm{R}^{(j)}_{2\pi/L_x} \bm{h}_{jk} \bm{R}^{(j)}_{-2\pi/L_x} & \text{if $\vec{r}_j-\vec{r}_k=(-1,0)$,}\\
        \bm{h}_{jk} & \text{otherwise.}
    \end{array}
    \right.
\end{align}
This modification is identical to that created by defect loops winding around the $y$-direction, as we now explain.

Following the construction of on-site symmetry defects specified in Ref.~\onlinecite{barkeshli14}, consider an $I_\theta$ defect loop that winds in the negative $y$-direction along the line $x=r^*_x-\frac{1}{2}$, where $r^*_x\in\{1, 2, \dots, L_x\}$. Let $C_\text{L}=\{j:[\vec{r}_j]_x=r^*_x\}$ be all the sites to the immediate left of $I_\theta$ and $C_\text{R}=\{j:[\vec{r}_j]_x=r^*_x-1\}$ be all the sites to the immediate right. For the nearest neighbor Hamiltonian with on-site $\textrm{U}(1)$ symmetry assumed in this section, $I_\theta$ can be created by the modification:
\begin{align}
    \bm{h}_j&\rightarrow\bm{h}_j\\
    \bm{h}_{jk}&\rightarrow\left\{
    \begin{array}{rl}
        \bm{R}^{(k)}_\theta \bm{h}_{jk} \bm{R}^{(k)}_{-\theta} & \text{if $[\vec{r}_j]_x=r^*_x$, $[\vec{r}_k]_x=r^*_x-1$,}\\
        \bm{R}^{(j)}_\theta \bm{h}_{jk} \bm{R}^{(j)}_{-\theta} & \text{if $[\vec{r}_j]_x=r^*_x-1$, $[\vec{r}_k]_x=r^*_x$,}\\
        \bm{h}_{jk} & \text{otherwise.}
    \end{array}
    \right.
\end{align}
By comparing this modification with Eq.~(\ref{eq:on-sitetrans}) and Eq.~(\ref{eq:nntrans}), we see that $\bm{H}(2\pi)$ is essentially $\bm{H}(0)$ with $I_{2\pi/L_x}$ defect loops wrapping the torus in the negative $y$-direction along the lines $x=\frac{1}{2}$, $x=\frac{3}{2}$, \dots, $x=L_x-\frac{1}{2}$. Since $|\Psi(2\pi)\rangle$ is a ground state of $\bm{H}(2\pi)$, it is essentially $|\Psi(0)\rangle$ with these defect loops.

Finally, we turn to $\bm{\mathcal{G}}|\Psi(2\pi)\rangle$. Rewriting the large gauge transformation as
\begin{align}
    \bm{\mathcal{G}}=\prod_{j:[\vec{r}_j]_x=1} \bm{R}^{(j)}_{2\pi /L_x} &\prod_{j:[\vec{r}_j]_x=2} \bm{R}^{(j)}_{4\pi/L_x} \nonumber \\
    &\dots\prod_{j:[\vec{r}_j]_x=L_x-1} \bm{R}^{(j)}_{2\pi-2\pi /L_x},
\end{align}
and recalling that defects lines obey the fusion rules
\begin{equation}
    I_{\theta_1}\times I_{\theta_2}=\coho{w}(\theta_1,\theta_2)I_{[\theta_1+\theta_2]_{2\pi}},
\end{equation}
where $\coho{w}(\theta_1,\theta_2)$ describes the $\textrm{U}(1)$ fractionalization, we can study the action of $\bm{\mathcal{G}}$ on $|\Psi(2\pi)\rangle$ step by step.
\begin{description}
    \item[Step 1] Applying $\prod_{j:[\vec{r}_j]_x=1} \bm{R}^{(j)}_{2\pi /L_x}$ moves $I_{2\pi/L_x}$ at $x=\frac{1}{2}$ to $x=\frac{3}{2}$, where it can be fused with $I_{2\pi/L_x}$ already there to form $I_{4\pi/L_x}$.
    \item[Step 2] Applying $\prod_{j:[\vec{r}_j]_x=2} \bm{R}^{(j)}_{4\pi/L_x}$ moves $I_{4\pi/L_x}$ at $x=\frac{3}{2}$ to $x=\frac{5}{2}$, where it can be fused with $I_{2\pi/L_x}$ already there to form $I_{6\pi/L_x}$.
    \begin{center}$\vdots$\end{center}
    \item[Step $\bm{L_x -1}$] Applying $\prod_{j:[\vec{r}_j]_x=L_x-1} \bm{R}^{(j)}_{2\pi-2\pi /L_x}$ moves $I_{2\pi-2\pi /L_x}$ at $x=L_x-\frac{3}{2}$ to $x=L_x-\frac{1}{2}$, where it can be fused with $I_{2\pi/L_x}$ already there to form a $v$ anyon loop along the negative $y$ direction, where $v=\coho{w}(\theta,2\pi-\theta)$.
\end{description}
Since $|\Psi(2\pi)\rangle$ is $|\Psi(0)\rangle$ with defect loops wrapping around the torus in the $y$-direction, and $\bm{\mathcal{G}}|\Psi(2\pi)\rangle$ is $|\Psi(2\pi)\rangle$ with all these defects loops fused to form a single $v$ anyon loop around the $y$-direction of the torus, we conclude that $\bm{\mathcal{G}}|\Psi(2\pi)\rangle=\bm{W}_v|\Psi(0)\rangle$.

\bibliographystyle{apsrev4-1}
\bibliography{Luttinger}

\end{document}